\documentclass[aps,prb,reprint,superscriptaddress,showpacs]{revtex4-1}

\usepackage[dvipdfmx]{graphicx}
\usepackage{bm}
\usepackage{color}
\usepackage{amsmath,amssymb}

\newcommand{\rr}{{\bf r}}
\newcommand{\q}{{\bf q}}

\begin{document}

\title{Odd-parity superconductivity by competing spin-orbit coupling and orbital effect \\ in artificial heterostructures}

\author{Tatsuya Watanabe}
\affiliation{Graduate School of Science and Technology, Niigata University, Niigata 950-2181, Japan}
\author{Tomohiro Yoshida}
\affiliation{Department of Physics, Gakushuin University, Tokyo 171-8588, Japan}
\author{Youichi Yanase}
\email[]{yanase@scphys.kyoto-u.ac.jp}
\affiliation{Department of Physics, Graduate School of Science, Kyoto University, Kyoto 606-8502, Japan}

\date{\today}

\begin{abstract}
We show that odd-parity superconductivity occurs in multilayer Rashba systems without requiring spin-triplet 
Cooper pairs.  A pairing interaction in the spin-singlet channel stabilizes the odd-parity 
pair-density-wave (PDW) state 
in the magnetic field parallel to the two-dimensional conducting plane. 
It is shown that the layer-dependent Rashba spin-orbit coupling and the orbital effect 
play essential roles for the PDW state in binary and tricolor heterostructures. 
We demonstrate that the odd-parity PDW state is a symmetry-protected topological superconducting state 
characterized by the one-dimensional winding number in the symmetry class {\it BDI}. 
The superconductivity in the artificial heavy-fermion superlattice CeCoIn$_5$/YbCoIn$_5$ and bilayer interface 
SrTiO$_3$/LaAlO$_3$ is discussed. 
\end{abstract}


\maketitle

\section{Introduction}

Parity is an essential quantum number of quantum phases unless inversion symmetry is broken. 
Classification of superconducting states is based on the parity of the order parameter.~\cite{Sigrist-Ueda} 
According to the conventional understanding,~\cite{Sigrist-Ueda} even-parity superconductivity is realized 
by the condensation of spin-singlet Cooper pairs, while odd-parity superconductivity is induced 
by spin-triplet Cooper pairs, because of the anticommutation relation of fermions. 
Even-parity superconductivity has been observed in a variety of materials, e.g., 
archetypal strongly correlated electron systems such as high-$T_{\rm c}$ cuprate 
superconductors (SCs)~\cite{Scalapino,Moriya-Ueda} as well as conventional SCs 
stabilized by electron-phonon coupling. 
On the other hand, only a few materials are considered as possible hosts of odd-parity superconductivity. 
This is probably because the conditions for spin-triplet pairing are unfavorable in most materials. 
Since electron-phonon coupling mostly stabilizes spin-singlet $s$-wave superconductivity, 
strong electron correlation is required for the glues of spin-triplet Cooper pairs. However, 
$d$-wave superconductivity is stable in most strongly correlated electron 
systems.~\cite{Yanase_review}

Odd-parity superconductivity has been attracting attention because of its multicomponent order parameters 
that give rise to multiple superconducting/superfluid phases and intriguing phenomena related to spontaneous 
symmetry breaking.~\cite{Sigrist-Ueda,Leggett,Sauls,Joynt,Maeno_review,Yanase_review2} 
Furthermore, a great deal of attention has recently been paid to odd-parity SCs because they are candidates 
for topological superconductivity.~\cite{Read-Green,Ivanov,Kitaev2001,Schnyder,Kitaev2009,Sato2010} 
However, only Sr$_2$RuO$_4$~\cite{Maeno_review,Yanase_review2} and some uranium-based heavy-fermion compounds such as 
UPt$_3$~\cite{Sauls,Joynt}, UGe$_2$~\cite{Saxena}, URhGe, and UCoGe~\cite{Aoki_review,Hattori} show 
strong evidence for spin-triplet pairing.

Recent theoretical studies have presented another way to stabilize the odd-parity superconducting state. 
It has been shown that odd-parity superconductivity may occur through spin-singlet Cooper pairs 
in crystals lacking local inversion symmetry. 
Such locally noncentrosymmetric crystals have a sublattice degree of freedom in electronic structures, 
allowing the odd-parity spin-singlet superconducting state.~\cite{Fischer,JPSJ.81.034702}  
Although this state is not allowed in the absence of spin-orbit coupling 
according to the BCS theory, such an exotic superconducting state may be stabilized by the sublattice-dependent 
spin-orbit coupling arising from the relativistic effect.~\cite{JPSJ.81.034702}
It has been shown that a long-range Coulomb interaction stabilizes odd-parity superconductivity 
in combination with spin-orbit coupling.~\cite{Fu-Berg,Nakosai} 
On the other hand, two of the authors have shown that the odd-parity spin-singlet superconducting state is stabilized 
by spin-orbit coupling and the paramagnetic effect without relying on the particular 
electron correlation effect.~\cite{PhysRevB.86.134514} 
Therefore, conventional electron-phonon coupling or antiferromagnetic spin fluctuation 
leading to spin-singlet Cooper pairing may induce odd-parity superconductivity 
when both spin-orbit coupling and the paramagnetic effect play important roles.

\begin{figure}[htbp]
\begin{center}
\includegraphics[width=75mm]{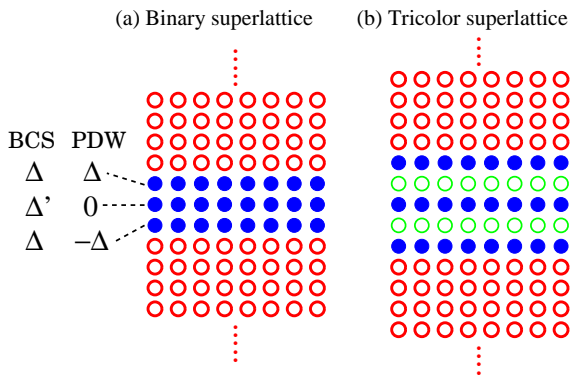}
\caption{(Color online) Illustration of superlattice structures studied in this paper. 
We show examples of the trilayer system ($M=3$). 
(a) Binary superlattice that has been realized in CeCoIn$_5$/YbCoIn$_5$. 
Closed (blue) and open (red) circles show the superconducting layers and spacer layers, respectively. 
The layer-dependent order parameters in the BCS and PDW states are described on the left. 
(c) Tricolor superlattice. Thick (red) and thin (green) open circles show the two kinds of spacer layers. 
}
\label{fig1}
\end{center}
\end{figure}

In a previous study, we focused on two-dimensional (2D) multilayer SCs, 
in which global inversion symmetry is preserved but some of the layers lack local inversion symmetry. 
Then, by applying a magnetic field along the {\it c} axis, the order parameter of spin-singlet superconductivity 
changes sign across the center layer, as shown in Fig.~1(a).~\cite{PhysRevB.86.134514} 
The order parameter spatially modulated in the atomic length scale ensures the odd parity of superconductivity. 
Such a superconducting state is called the pair-density-wave (PDW) state. 
Interestingly, the PDW state is classified into topological crystalline superconductivity 
protected by mirror reflection symmetry when the number of superconducting layers is odd.~\cite{Yoshida2015} 
A non-trivial topological invariant in the symmetry class {\it D}, the mirror Chern number, 
ensures the appearance of the Majorana edge mode at the edge. 
A promising candidate for realizing such a topological superconducting state 
is the recently grown artificial superlattices CeCoIn$_5$/YbCoIn$_5$ composed of the quasi-2D 
heavy-fermion SC CeCoIn$_5$ and the conventional metal YbCoIn$_5$.~\cite{NatPhys.7.849} 
The superconductivity occurs in CeCoIn$_5$ multilayers, and YbCoIn$_5$ plays the role of 
spacer layers. Thus, the superlattice is regarded as a 2D multilayer SC when the number of 
YbCoIn$_5$ layers is large. 
The strong spin-orbit coupling and the large paramagnetic effect~\cite{Izawa2001,Tayama2002} in CeCoIn$_5$ 
are favorable for the topological odd-parity superconductivity discussed above. 
Indeed, the effects of layer-dependent Rashba spin-orbit coupling (RSOC) on the superconducting state 
of CeCoIn$_5$/YbCoIn$_5$ have been observed.~\cite{Goh,shimozawa}

On the other hand, assuming a large Maki parameter $\alpha_{\rm M} = \sqrt{2}H_{\rm c2}^{\rm orb}/H_{\rm c2}^{\rm P}$, 
Ref.~23 neglected the orbital effect that competes with the paramagnetic effect. 
$H_{\rm c2}^{\rm orb}$ and $H_{\rm c2}^{\rm P}$ are fictitious upper critical fields determined 
by the orbital effect (orbital limit) and 
by the paramagnetic effect (paramagnetic limit), respectively. 
Since the Maki parameter of bulk CeCoIn$_5$ is moderate ($\alpha_{\rm M} \sim 3$) for the magnetic field 
along the {\it c} axis,~\cite{Miclea} the orbital effect may suppress the PDW state. 
A simple way to reduce the orbital effect is to apply a magnetic field along the 2D conducting plane. 
However, when the orbital effect is completely neglected, the in-plane magnetic field induces the complex stripe 
(CS) state,~\cite{Yoshida2013} and the PDW state is not stabilized. 
In this paper, we show that the odd-parity PDW state is stabilized by switching on a weak orbital effect.

Furthermore we find that the PDW state is a topological crystalline SC protected by magnetic mirror symmetry. 
Because mirror reflection symmetry with respect to the {\it ab} plane is broken by the in-plane magnetic field, 
the mirror Chern number is no longer a topological invariant. However, magnetic mirror symmetry is preserved 
when we apply the magnetic field along the {\it a} axis or the {\it b} axis. Using this symmetry, we define 
the topological invariant, i.e., the one-dimensional winding number in the symmetry class {\it BDI}. 
We will show a non-trivial winding number and the resulting Majorana edge state. 

The organization of this paper is as follows. 
In Sec.~II we introduce the model for multilayer SCs possessing the layer-dependent RSOC. 
Various superconducting states obtained by solving the linearized mean-field equation are illustrated. 
We study the binary superlattices in Sec.~III and discuss the superconducting state in the artificial 
superlattice CeCoIn$_5$/YbCoIn$_5$. The tricolor superlattice in Fig.~1(b) is also investigated in Sec.~IV. 
In Secs.~III and IV, it is shown that the odd-parity PDW state is stabilized through the RSOC and the orbital effect. 
In Sec.~V, the topologically non-trivial properties of the PDW state are clarified, and the Majorana edge state is demonstrated. 
A brief summary and discussions are provided in Sec.~VI.

\section{Model and Formulation}

\subsection{Model}

First, we introduce the model for 2D multilayer SCs. 
By simply neglecting the spacer layers, the binary and tricolor superlattices in Fig.~1 
are described by the multilayer model. 
By taking into account the layer-dependent RSOC, the orbital effect, and Zeeman coupling, 
the Hamiltonian is described as
\begin{eqnarray}
{\cal H}&=& \sum_{{\bm k},s,m} \xi({\bm k}+{\bm p}_m) c_{{\bm k}sm}^\dagger c_{{\bm k}sm} 
+t_\perp \sum_{{\bm k},s,\langle m,m'\rangle} c^\dagger_{{\bm k}sm}c_{{\bm k}sm'} 
\nonumber \\
&&+ \sum_{{\bm k},{\bm k}',{\bm q},m} V({\bm k},{\bm k}') c_{{\bm k_+} \uparrow m}^\dagger 
c_{-{\bm k_-} \downarrow m}^\dagger c_{-{\bm k'_-} \downarrow  m} c_{{\bm k'_+} \uparrow  m}  \nonumber \\ 
&& - \sum_{{\bm k},s,s',m} \mu_{\rm B} {\bm H} \cdot {\bm \sigma}_{ss'} c_{{\bm k}sm}^\dagger c_{{\bm k}s'm} \nonumber \\ 
&&+\sum_{{\bm k},s,s',m} \alpha_m {\bm g}({\bm k}+{\bm p}_m) \cdot {\bm \sigma}_{ss'}c^\dagger_{{\bm k}sm}c_{{\bm k}s'm},  
\label{eq:model}
\end{eqnarray}
where ${\bm k}$, $s$, and $m$ $(= 1, . . . ,M)$ are indexes of momentum, spin, and layer, respectively. 
The number of superconducting layers is $M$.

The first term is the energy dispersion in the single-layer limit. 
We adopt the nearest neighbor hopping term in the square lattice for simplicity, 
\begin{equation}
\xi({\bm k}) = -2 t (\cos{k_xa}+\cos{k_ya}) - \mu,  
\end{equation}
where $a$ is the lattice constant. 
The orbital effect induced by the applied magnetic field is taken into account through the Peierls phase. 
When we consider the magnetic field along the [100]-axis, ${\bm H} = H\hat{x}$, we can choose the vector 
potential ${\bm A} = (0, -Hz, 0)$. Then, the orbital effect leads to 
the layer-dependent shift of momentum, ${\bm k} \rightarrow {\bm k} + \frac{e}{ \hbar} {\bm A}$.  
Thus, we obtain ${\bm p}_m = \frac{e}{ \hbar} H d \left[ m-(M+1)/2 \right] \hat{y}$, 
where $d$ is the lattice spacing between the nearest neighbor superconducting layers. 
For binary superlattices, $d=c$ with $c$ being the lattice constant along the {\it c}-axis. 
Later we adopt the lattice constant of CeCoIn$_5$, $a=4.6$ \AA $\,$ and $c=7.5$ \AA $\,$
since we focus on the artificial superlattice CeCoIn$_5$/YbCoIn$_5$.~\cite{NatPhys.7.849} 
The second term of Eq.~(\ref{eq:model}) is the inter-layer hopping term. 
Since we consider the heterostructures composed of quasi-2D SCs, 
the inter-layer hopping $t_\perp$ is assumed to be much smaller than the in-plane hopping, $t_\perp \ll t$.

The third term describes the pairing interaction. We assume $s$-wave superconductivity for simplicity, and thus, 
\begin{equation}
V({\bm k},{\bm k}')=-V_{\rm s}. 
\end{equation} 
We have confirmed that qualitatively the same results are obtained for $d$-wave superconductivity. 
As we will show layer, a spatially non-uniform superconducting state may be stabilized. 
Thus, we take into account the finite center-of-mass momentum of Cooper pairs $\bm{q}$, 
and $\bm{k}_\pm = \bm{k}\pm \bm{q}/2$.
The fourth term is the Zeeman coupling term giving rise to the paramagnetic effect on the superconducting state. 

We show that exotic superconducting states are stabilized by the layer-dependent RSOC
represented in the last term of Eq.~(1), which arises from the local violation of inversion symmetry.~\cite{JPSJ.81.034702} 
Ensured by the global inversion symmetry in the crystal structure, the RSOC is odd with respect to the mirror reflection 
on the center layer, and thus $\alpha_{M+1-m} = -\alpha_m$. 
For example, the layer-dependent coupling constant is $(\alpha_1,\alpha_2)=(\alpha,-\alpha)$ for bilayers and 
$(\alpha_1,\alpha_2,\alpha_3)=(\alpha,0,-\alpha)$ for trilayers. 
We assume a g-vector characterizing the RSOC,~\cite{Rashba} 
${\bm g}({\bm k})= (-\sin k_ya, \sin k_xa, 0)$, 
so as to satisfy the periodicity in momentum space.

For bilayers, a similar model has been investigated by noticing the twin boundary of noncentrosymmetric 
SCs.~\cite{Iniotakis,Arahata,Achermann,Aoyama} Then, the intriguing superconducting phase with broken time-reversal 
symmetry was investigated by assuming comparable pairing interactions 
for the spin-singlet Cooper pairs and spin-triplet ones.~\cite{Iniotakis,Arahata,Achermann} 
We avoid such fine-tuning of pairing interactions here 
and consider the dominantly spin-singlet pairing state, which is realized in most SCs. 
Aoyama {\it et. al.} studied the magneto-electric effect on the upper and lower critical magnetic fields,~\cite{Aoyama} 
but their Ginzburg-Landau model does not appropriately take into account the paramagnetic effect in the high magnetic field 
region, and therefore, the PDW state focused in this paper is not obtained.

We assume small inter-layer hopping $t_\perp/t=0.1$ and a moderate RSOC $\alpha/t=0.3$, 
unless explicitly mentioned otherwise. 
As shown by previous studies,~\cite{JPSJ.81.034702,PhysRevB.86.134514,Yoshida2013} 
exotic superconducting states may be stabilized when $|\alpha|/t_\perp \gtrsim 1$. 
Thus, we assume $|\alpha|/t_\perp \gtrsim 1$ throughout this paper. 
This condition may be satisfied in heterostructures of quasi-2D compounds. 
The chemical potential is $\mu/t=2$ in Sec~III and IV while it is $\mu/t=-2$ in Sec~V. 
We choose the pairing interaction $V_{\rm s}/t=1.3$ or $V_{\rm s}/t=1.5$. 
We confirmed that the following results are almost independent of the choice of $V_{\rm s}/t$.

\subsection{Linearized mean-field theory}

We study the superconducting state by means of the linearized mean-field theory. 
Although we have to fully solve the Bogoliubov-de Gennes (BdG) equation in order to obtain the superconducting 
phase diagram, we can clarify the superconducting state near the transition temperature 
by linearizing the BdG equation while avoiding the numerical limitations of the full BdG equation. 
The linearized BdG equation is formulated by calculating the superconducting susceptibility, 
\begin{equation}
\chi^{\rm sc}_{m m'}(q) = \int_0^\beta d\tau e^{i \Omega_n \tau} \langle B_{{\bm q} m}(\tau) B^{\dagger}_{{\bm q} m'}(0)  \rangle, 
\end{equation}
where $q=({\bm q},i\Omega_n)$ and $\Omega_n = 2 \pi n k_{\rm B} T$ is the boson Matsubara frequency. 
The annihilation operator of Cooper pairs is introduced as 
\begin{equation}
B_{{\bm q} m}= \sum_{\bm k} c_{{\bm k}+{\bm q} \uparrow m} c_{-{\bm k} \downarrow m},  
\end{equation}
and
$B_{{\bm q} m}(\tau)= e^{-{\cal H} \tau}B_{{\bm q} m}e^{{\cal H} \tau} $. 

The superconducting susceptibility is obtained by using the T-matrix approximation, 
\begin{equation}
\hat{\chi}^{\rm sc}(q) = \frac{\hat{\chi}^0(q)}{\hat{1}-V_{\rm s}\hat{\chi}^0(q)}, 
\end{equation}
where $\hat{\chi}^{\rm sc} = \left(\chi^{\rm sc}_{mm'}\right)$ is the $M \times M$ matrix.
The irreducible susceptibility is calculated by  
\begin{eqnarray}
\chi_{m m'}^0(q) &=&\frac{1}{\beta} \sum_{{\bm k},l} 
\bigl[G_{m m'}^{\uparrow \uparrow}({\bm k}+{\bm q},i\omega_l) G_{m m'}^{\downarrow \downarrow}(-{\bm k},i\Omega_n - i\omega_l) \nonumber \\
&&-G_{m m'}^{\uparrow \downarrow}({\bm k}+{\bm q},i\omega_l) G_{m m'}^{\downarrow \uparrow}(-{\bm k},i\Omega_n - i\omega_l) \bigr], 
\label{chi}
\end{eqnarray}
where $G_{m m'}^{s s'}({\bm k},i\omega_l)$ is the non-interacting Green function and $\omega_l=(2l+1)\pi k_{\rm B} T$ 
is the fermion Matsubara frequency. 

Superconducting transition occurs at the temperature where $\hat{\chi}^{\rm sc}(q)$ diverges. 
Thus, the criterion of the superconducting instability is obtained as the largest eigenvalue of 
$V_{\rm s}\hat{\chi}^0(q)$ is unity. 
We obtain the layer-dependent order parameter $\Delta_m(\rr)=\Delta_m e^{i\q \cdot \rr}$ from the eigenvector, 
$(\Delta_1, \Delta_2, ... \Delta_M)^{\rm T}$. 
Because the global inversion symmetry is preserved in our model, the eigenvalues are equivalent 
between the momentum $\pm\q$. 
Hence, the single-$q$ state or the double-$q$ state may be stabilized when the center-of-mass 
momentum of Cooper pairs is finite. 
It is expected that the double-$q$ state is stable in our model because the order parameter 
almost disappears in one of the outermost layers in the single-$q$ state with a small condensation energy. 
For instance, $\Delta_M \ll \Delta_1$ for momentum ${\bm q}$ while $\Delta_1 \ll \Delta_M$ 
for the opposite momentum $-{\bm q}$. 
In the double-$q$ state, the order parameter is described as 
$\Delta_m(\rr)=\Delta_m^{(+)} e^{i\q \cdot \rr} + \Delta_m^{(-)} e^{-i\q \cdot \rr}$, 
where $\Delta_m^{(\pm)}$ is the eigenvector of $V_{\rm s}\hat{\chi}^0(q)$ for the momentum $\pm \q$, respectively. 
We confirmed that the bosonic Matsubara frequency is always zero, $\Omega_n =0$.

\subsection{Superconducting states}

\begin{figure}[htbp]
\begin{center}
\includegraphics[width=85mm]{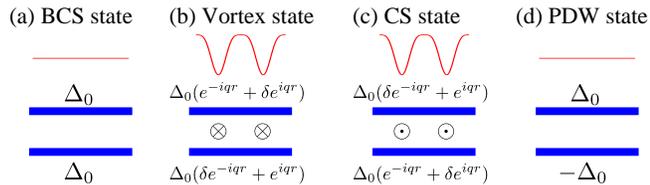}
\caption{(Color online) Illustration of superconducting states in the bilayer system. 
(a) BCS state, (b) vortex state, (c) CS state, and (d) PDW state. 
Thick bars show the superconducting layers.
Layer-dependent order parameters are described in the figures. 
The spatial dependence of the superconducting gap is illustrated by thin lines on top of the figures. 
The vortex in (b) and anti-vortex in (c) are shown by arrows. 
}
\label{fig2}
\end{center}
\end{figure}

In this subsection, we classify the solution of the linearized BdG equation. 
As we will show later, various superconducting states are stabilized in our model. 
They are illustrated for the bilayer system in Fig.~\ref{fig2}. 
In this subsection we discuss the bilayer system for simplicity, since the extension to more-than-two-layer systems 
is straightforward.

The uniform superconducting state [$\Delta_m(\rr) = \Delta$] is stable at zero magnetic field 
as expected from conventional BCS theory. 
Thus, we call the uniform state the ``BCS state'' [Fig.~\ref{fig2}(a)]. 
On the other hand, a variety of spatially non-uniform states may be stabilized in the magnetic field. 
First, the orbital effect induces the vortex state illustrated in Fig.~\ref{fig2}(b). 
When the quantum vortices penetrate inside multilayers, the order parameter is described as 
\begin{eqnarray}
&& \Delta_1({\bf r}) = \Delta_0 (e^{-i{\bf q} \cdot {\bf r}}+\delta e^{i{\bf q} \cdot {\bf r}}), 
\label{vs1}\\
&& \Delta_2({\bf r}) = \Delta_0 (\delta e^{-i{\bf q} \cdot {\bf r}}+e^{i{\bf q} \cdot {\bf r}}),   
\label{vs2}
\end{eqnarray}
where $|\delta| < 1$, and $\q = q (\hat{z} \times \hat{H})$ with $\hat{H} = {\bm H}/|{\bm H}|$ and $q>0$. 
Second, the layer-dependent RSOC stabilizes the CS state through the paramagnetic effect.~\cite{Yoshida2013} 
The CS state is also described by Eqs.~(\ref{vs1}) and (\ref{vs2}), 
however, the sign of the center-of-mass momentum $q$ 
depends on the band structure and the sign of RSOC. 
For our choice of parameters, the RSOC favors $q < 0$ when $\alpha > 0$ while $q > 0$ when $\alpha <0$. 
Thus, the CS state is regarded as an anti-vortex state [Fig.~\ref{fig2}(c)] and is distinguished 
from the vortex state, when $\alpha > 0$. 
Then, the spin-orbit coupling competes with the orbital effect, and gives rise to an intriguing superconducting 
phase diagram, as we show later. 
We focus on this case in the following, although the other case, $\alpha <0$, is briefly discussed.

Finally, Fig.~\ref{fig2}(d) illustrates the PDW state, where 
$\left[\Delta_1(\rr), \Delta_2(\rr)\right] = \left(\Delta, -\Delta\right)$.~\cite{PhysRevB.86.134514} 
The order parameter is uniform in the 2D conducting plane, but it changes sign between layers. 
As we mentioned before, the PDW state is an odd-parity superconducting state, 
although the superconductivity is induced by the spin-singlet $s$-wave Cooper pairs. 
Although we have shown that the PDW state is stabilized in the {\it c}-axis magnetic field 
near the Pauli limit,~\cite{PhysRevB.86.134514} 
in this paper we show that the PDW state is also stabilized in the in-plane magnetic field when the RSOC competes 
with the orbital effect.

\begin{table}[ht]
\begin{center}
\begin{tabular}{c||l|l}
 & Bilayer & Trilayer \\ \hline\hline
BCS & $\Delta_1({\bf r}) = \Delta$  & $\Delta_1({\bf r}) = \Delta$ \\ 
    & $\Delta_2({\bf r}) = \Delta$  & $\Delta_2({\bf r}) = \Delta'$ \\ 
    &                                 & $\Delta_3({\bf r}) = \Delta$ \\ \hline
PDW & $\Delta_1({\bf r}) = \Delta$  & $\Delta_1({\bf r}) = \Delta$ \\ 
    & $\Delta_2({\bf r}) = -\Delta$ & $\Delta_2({\bf r}) = 0$ \\ 
    &                                 & $\Delta_3({\bf r}) = -\Delta$ \\ \hline
FFLO & $\Delta_1({\bf r})  = \Delta \cos({\bf q} \cdot {\bf r})$ &  $\Delta_1({\bf r})  = \Delta \cos({\bf q} \cdot {\bf r})$ 
\\    & $\Delta_2({\bf r})  = \Delta \cos({\bf q} \cdot {\bf r})$ & 
$\Delta_2({\bf r})  = \Delta' \cos({\bf q} \cdot {\bf r})$ \\ 
     &                                                                      &    
$\Delta_3({\bf r})  = \Delta \cos({\bf q} \cdot {\bf r})$ \\ \hline
Vortex & $\Delta_1({\bf r}) = \Delta (e^{-i{\bf q} \cdot {\bf r}}+\delta e^{i{\bf q} \cdot {\bf r}})$ & 
$\Delta_1({\bf r}) = \Delta (e^{-i{\bf q} \cdot {\bf r}}+\delta e^{i{\bf q} \cdot {\bf r}})$ \\ 
    & $\Delta_2({\bf r}) = \Delta (\delta e^{-i{\bf q} \cdot {\bf r}}+e^{i{\bf q} \cdot {\bf r}})$  & 
$\Delta_2({\bf r}) = \Delta' \cos({\bf q} \cdot {\bf r})$ \\ 
    &                                                                                     &
$\Delta_3({\bf r}) = \Delta (\delta e^{-i{\bf q} \cdot {\bf r}}+e^{i{\bf q} \cdot {\bf r}})$ \\ \hline

CS & $\Delta_1({\bf r}) = \Delta (e^{i{\bf q} \cdot {\bf r}}+\delta e^{-i{\bf q} \cdot {\bf r}})$ & 
$\Delta_1({\bf r}) = \Delta (e^{i{\bf q} \cdot {\bf r}}+\delta e^{-i{\bf q} \cdot {\bf r}})$ \\ 
    & $\Delta_2({\bf r}) = \Delta (\delta e^{i{\bf q} \cdot {\bf r}}+e^{-i{\bf q} \cdot {\bf r}})$  & 
$\Delta_2({\bf r}) = \Delta' \cos({\bf q} \cdot {\bf r})$ \\ 
    &                                                                                     &
$\Delta_3({\bf r}) = \Delta (\delta e^{i{\bf q} \cdot {\bf r}}+e^{-i{\bf q} \cdot {\bf r}})$ \\ 
\end{tabular}
\caption{Layer-dependent order parameter of the BCS, PDW, FFLO, vortex, and CS states in bilayers and trilayers.}
\label{tab:orderpara}
\end{center}
\end{table}

The BCS, PDW, vortex, and CS states are distinguished from each other in more-than-two-layer systems too. 
In Table~I, we summarize the order parameter of these states in trilayers as well as in bilayers. 
The order parameter of the {\it so-called} Fulde-Ferrell-Larkin-Ovchinnikov (FFLO) state~\cite{FF,LO} is also shown for a comparison.

\section{Binary Superlattice}

\subsection{Bilayer system}

In this section, we study the binary superlattices (see Fig.~\ref{fig1}(a)) 
which have been fabricated in CeCoIn$_5$/YbCoIn$_5$.~\cite{NatPhys.7.849}

We begin with the simplest case, namely, the bilayer system ($M=2$) illustrated in Fig.~\ref{fig2}. 
Then, the strength of the orbital effect is controlled by the Fermi energy $E_{\rm F}$, 
which is proportional to the in-plane hopping $t$. 
The orbital effect is estimated from the dimensionless quantity $H \xi c/\Phi_0$  with $\Phi_0 = \frac{h}{2e}$ 
being the flux quantum. Since the coherence length of superconductivity is 
$\xi \simeq \hbar v_{\rm F}/k_{\rm B}T_{\rm c} \sim \left(E_{\rm F}/k_{\rm B} T_{\rm c}\right) a$,  
the orbital effect is enhanced by increasing the Fermi energy. On the other hand, the paramagnetic effect of 
the magnetic field is estimated from another dimensionless quantity $\mu_{\rm B} H/k_{\rm B}T_{\rm c}$, 
that is independent of the Fermi energy. 
Thus, the orbital effect (paramagnetic effect and RSOC) plays an important role in the superconducting state 
for large (small) in-plane hopping $t$. Note that the RSOC induces the CS state through the paramagnetic effect.

\begin{figure}[h]
  \begin{center}
\includegraphics[width=80mm]{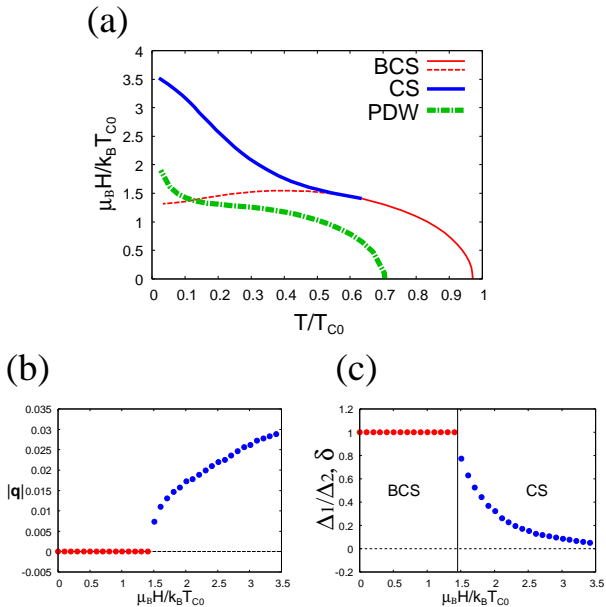}
    \caption{
(Color online) (a) Transition temperatures of various superconducting states 
in the bilayer system. We assume a small in-plane hopping, $t=20$ meV. Thin (red), moderate (blue), and thick (green) 
lines show the $T_{\rm c}$ of BCS, CS, and PDW states, respectively. 
The highest transition temperature is observable and drawn by the solid line, while the dashed and dot-dashed lines 
show the fictitious transition temperatures. 
(b) Magnetic field dependence of the center-of-mass momentum $|\q|$. 
(c) Layer dependence of the order parameter. 
We show $\Delta_2/\Delta_1$ for the BCS state and $\delta$ for the CS state. 
In this subsection, we choose the pairing interaction $V_{\rm s}/t=1.5$, which gives the transition temperature 
$k_{\rm B} T_{\rm c0} = 0.0124 t$ in the absence of the spin-orbit coupling and magnetic field.  
The temperature and magnetic field are scaled by $T_{\rm c0}$ in the figures. 
}
    \label{fig3}
  \end{center}
\end{figure}

When we assume small in-plane hopping $t=20$ meV consistent with the heavy effective mass of CeCoIn$_5$, 
the orbital effect is negligible. Indeed, we obtain the phase diagram in Fig.~\ref{fig3}(a) 
which resembles to the result in the paramagnetic limit.~\cite{Yoshida2013}  
The CS state is stable in the high magnetic field region, but the PDW state is not stabilized. 
As increasing the magnetic field, the center-of-mass momentum of Cooper pairs gradually increases through the 
BCS-CS phase transition (Fig.~\ref{fig3}(b)), indicating the second order phase transition. 
Figure~\ref{fig3}(c) shows that $\Delta_1 = \Delta_2$ in the BCS state while 
$\delta$ in Eqs.~(\ref{vs1}) and (\ref{vs2}) decreases with increasing the magnetic field in the CS state. 
These behaviors are consistent with the previous study on the same model~\cite{Yoshida2013} 
where the orbital effect is simply neglected. 
Thus, the previous study that was focused on the heavy-fermion superlattice CeCoIn$_5$/YbCoIn$_5$ is justified.

\begin{figure}[htbp]
\begin{center}
\includegraphics[width=80mm]{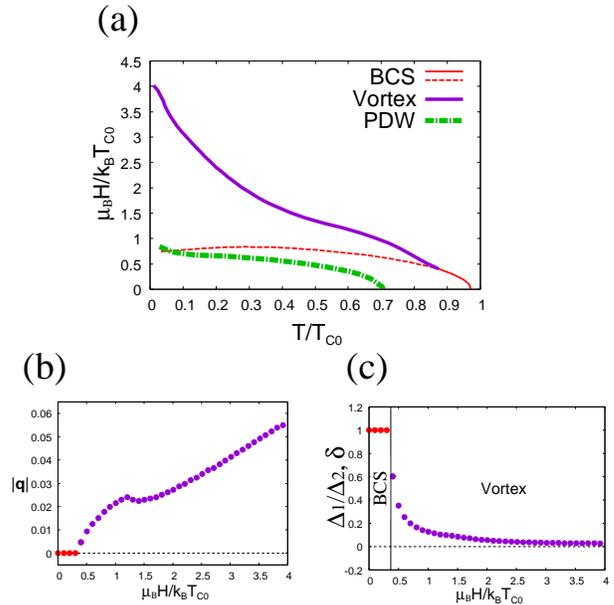}
\caption{(Color online) Transition temperatures (a), center-of-mass momentum (b), and layer-dependent order parameter (c) 
for a large in-plane hopping, $t=200$ meV. The other parameters are the same as those in Fig.~\ref{fig3}. 
The moderate (purple) line in (a) shows the transition temperature of the vortex state. 
} 
\label{fig4}
\end{center}
\end{figure}

On the other hand, the orbital effect significantly affects the superconducting state for large in-plane hopping, 
$t=200$ meV. 
Figure~\ref{fig4}(a) shows that the vortex state is stable in the high magnetic field region, as expected. 
The BCS-vortex phase transition is second order as indicated by the continuous change of the center-of-mass momentum 
[Fig.~\ref{fig4}(b)] and $\delta$ [Fig.~\ref{fig4}(c)]. 
Although the RSOC and the paramagnetic effect play less important roles than the orbital effect, 
they induce a characteristic magnetic field dependence in the center-of-mass momentum around the Pauli-Chandrasekhar-Clogston 
limit $\mu_{\rm B}H/k_{\rm B}T_{\rm c0} = 1.25$. 
The RSOC and paramagnetic effect suppress the orbital effect, and thus decrease the center-of-mass momentum 
above the Pauli-Chandrasekhar-Clogston limit.

\begin{figure}[htbp]
\begin{center}
\includegraphics[width=65mm]{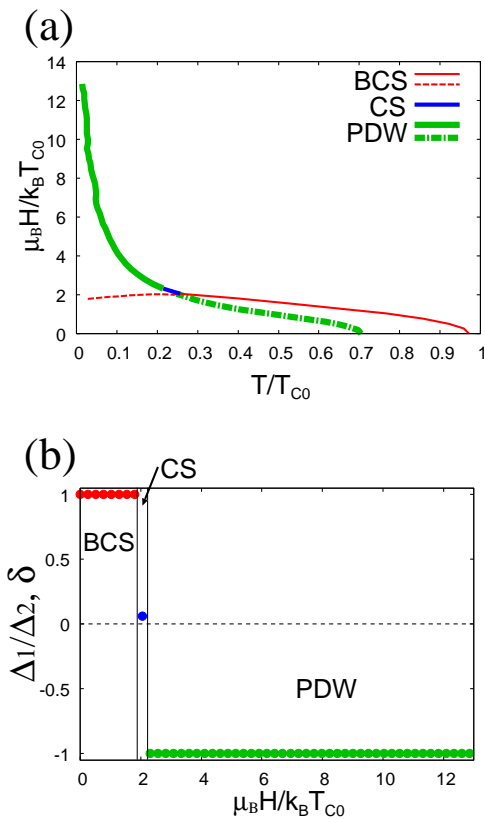}
\caption{(Color online) Transition temperatures (a) and layer dependence of the order parameter (b) 
for a moderate in-plane hopping, $t=80$ meV. The other parameters are the same as those in Fig.~\ref{fig3}. 
It is shown that the PDW state is stable in the high magnetic field region. 
} 
\label{fig5}
\end{center}
\end{figure}

A main result of our study is obtained when the RSOC competes with the orbital effect. 
Such a situation is realized for moderate in-plane hopping $t=80$ meV. 
Then, the PDW state is stabilized in the high magnetic field region, as shown in Fig.~\ref{fig5}. 
Note that the PDW state is induced neither by the orbital effect nor by the paramagnetic effect and RSOC. 
The PDW state is stable due to a balance of these effects. We explain this mechanism in details here. 
The order parameter of the CS and vortex states is described by Eqs.~(\ref{vs1}) and (\ref{vs2}), and  
these two states are differentiated by the sign of $q$. The positive $q$ (vortex state) is favored by the orbital effect, 
although the negative $q$ (CS state) is induced by the RSOC. 
Thus, $q \sim 0$ when these two effects are in balance. Then, the inter-layer Josephson coupling stabilizes 
the uniform superconducting state along the conducting plane, and thus $q=0$. 
The $\pi$ phase difference between layers 
is favored so that the paramagnetic depairing effect is avoided 
as in the $c$-axis magnetic field.~\cite{PhysRevB.86.134514}  
In this way, the PDW state is stabilized by the orbital effect, the paramagnetic effect, 
RSOC, and inter-layer coupling.

\subsection{More-than-two-layer system}

Although it is hard to experimentally control the Fermi energy, the number of superconducting layers $M$ can be tuned 
by using the artificial superlattice.~\cite{NatPhys.7.849,Goh,shimozawa} 
Thus, we may be able to control the orbital effect by tuning $M$.  
Since the shift of momentum on the outermost layers, $|p_1| = |p_M| = e Hc (M-1)/2\hbar$, increases with $M$,  
the orbital effect is enhanced by increasing the number of superconducting layers. 
This is reasonable because vortices easily penetrate inside of thick SCs. 
We demonstrate here that the competing region of the RSOC and the orbital effect is realized by tuning $M$, 
and then the PDW state is stabilized.

We take into account the RSOC on the outermost layers, $(\alpha_1, \alpha_M) = (\alpha,-\alpha)$, while 
the RSOC on the other layers is neglected for simplicity. 
This is a reasonable assumption for the layer dependence of RSOC, because the spin-orbit coupling is determined 
by the local environment of atoms,~\cite{Yanase2008} and thus the outermost layers contain the largest spin-orbit coupling. 

\begin{figure}[htbp]
\begin{center}
\includegraphics[width=60mm]{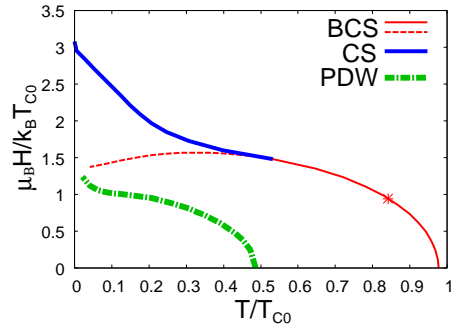}
\caption{(Color online) Transition temperatures of the BCS, CS, and PDW states in the trilayer system ($M=3$) 
for small in-plane hopping, $t=21$ meV. In this subsection, we assume $V_{\rm s}/t=1.3$, which gives rise to 
the transition temperature at zero magnetic field, $T_{\rm c0}=0.00487 t/k_{\rm B}$, in the absence of RSOC.
} 
\label{fig6}
\end{center}
\end{figure}

We fix the in-plane hopping, $t=21$ meV, and the pairing interaction, $V_{\rm s}/t=1.3$. 
Then, the orbital effect is negligible in the bilayer system as in Fig.~\ref{fig3}. 
The trilayer system shows the phase diagram (Fig.~\ref{fig6}) similar to Fig.~\ref{fig3}, 
and thus the trilayer system is still close to the Pauli limit.

\begin{figure}[htbp]
\begin{center}
\includegraphics[width=65mm]{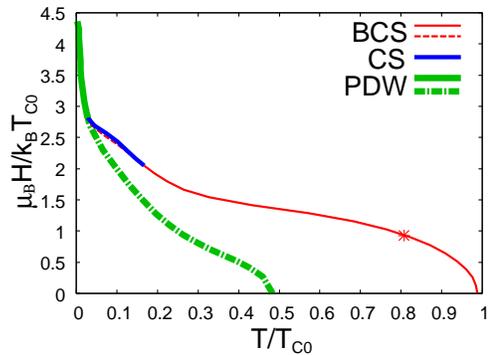}
\caption{(Color online) Transition temperatures of the BCS, CS, and PDW states in the five-layer system ($M=5$)
for small in-plane hopping, $t=21$ meV.
} 
\label{fig7}
\end{center}
\end{figure}

On the other hand, the orbital effect plays an important role in the five-layer system ($M=5$). 
Then, the orbital effect competes with the RSOC, and therefore, the PDW state is stabilized as expected from 
the results in Sec.~IIIA. Indeed, Fig.~\ref{fig7} shows that the BCS state, CS state, and PDW state are stabilized 
in the low, intermediate, and high magnetic field regions, respectively. 
Because the CS-PDW transition is first order phase transition, the upper critical field shows a kink, 
although the kink may be weak in some cases (see Figs.~5(a) and 10(a)). Generally speaking, a distinct kink appears 
when the transition temperature of the PDW state is small. Then, the $H_{\rm c2}$ of the CS state is suppressed, 
while that of the PDW state shows an upward curvature.  
The observation of the kink will be an experimental test for the presence of the PDW state.

\begin{figure}[htbp]
\begin{center}
\includegraphics[width=60mm]{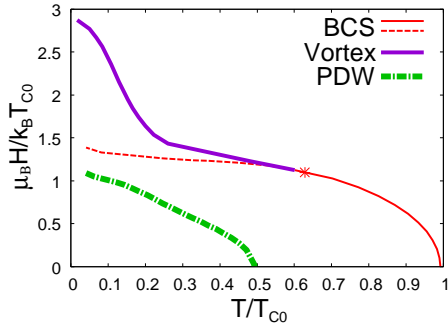}
\caption{(Color online) Transition temperatures of the BCS, vortex, and PDW states in the seven-layer system ($M=7$) 
for small in-plane hopping, $t=21$ meV. 
} 
\label{fig8}
\end{center}
\end{figure}

When we furthermore increase the number of superconducting layers, the superconducting state is dominated by the orbital effect, 
and thus the vortex state is stabilized in the high magnetic field region. For example, we show the phase diagram of 
the seven-layer system ($M=7$) in Fig.~\ref{fig8}. It is shown that the PDW state is not stabilized.

\begin{figure}[htbp]
\begin{center}
\includegraphics[width=84mm]{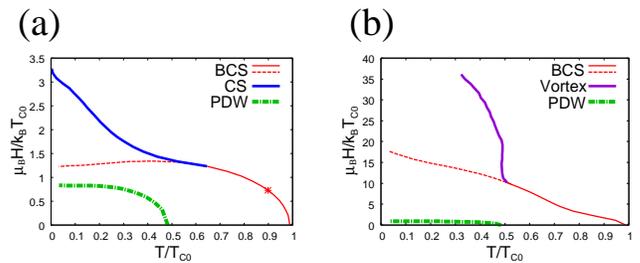}
\caption{(Color online) 
Transition temperatures of the superconducting states in the five-layer system. 
(a) Pauli limit by setting ${\bm p}_m =0$. 
(b) Orbital limit by eliminating the Zeeman term in the Hamiltonian [Eq.~(\ref{eq:model})]. 
The other parameters are the same as those in Fig.~\ref{fig7}. 
} 
\label{fig9}
\end{center}
\end{figure}

We focus here on the five-layer system, and we emphasize the cooperative role of the RSOC, the orbital effect, 
and the paramagnetic effect on the thermodynamical stability of the PDW state. 
Figure~\ref{fig9}(a) shows the phase diagram in the absence of the orbital effect. 
The CS state is more stable than the PDW state, as in the bilayer and trilayer systems. 
On the other hand, the vortex state is stable in the high magnetic field region 
when we neglect the paramagnetic effect, as shown in Fig.~\ref{fig9}(b). 
Note that the RSOC does not play important roles in the absence of the paramagnetic effect.

It should be noticed that the orbital limit of the upper critical field [Fig.~\ref{fig9}(b)] is much larger than the 
paramagnetic limit [see Fig.~\ref{fig9}(a)], indicating the large Maki parameter. 
This means that the upper critical field of five-layer systems is mainly determined from the paramagnetic effect. 
In this sense, the PDW state occurs near the paramagnetic limit, although a weak orbital effect is needed. 
Stars in Figs.~\ref{fig6}, \ref{fig7}, \ref{fig8}, and \ref{fig9}(a) show the crossover  
induced by the paramagnetic effect.~\cite{PhysRevB.86.134514} Because the center layer is not protected against 
the paramagnetic effect by the RSOC, the order parameter in the center layer $\Delta_{(M+1)/2}$ suddenly decreases 
by increasing the magnetic field through the crossover. We see that the paramagnetic effect appears even in the 
seven-layer system, where the orbital effect is larger than the effect of RSOC.

\subsection{CeCoIn$_5$/YbCoIn$_5$}

In the previous subsections, we designed the odd-parity PDW state using the artificial heterostructures. 
It has been shown that the orbital effect is controlled by the number of superconducting layers. 
Indeed, various superlattices CeCoIn$_5$/YbCoIn$_5$ with $M>2$ are superconducting, and we can tune the 
number of layers $M$.~\cite{NatPhys.7.849,Goh,shimozawa} Thus, the artificial superlattice CeCoIn$_5$/YbCoIn$_5$ 
may be a new platform for the odd-parity superconductivity. A reasonable parameter $t \sim 20$ meV 
leads to the PDW state in the five-layer system.

We comment here on the recent experimental observations of the paramagnetic effect 
in the superlattice CeCoIn$_5$/YbCoIn$_5$.~\cite{Goh} Goh {\it et. al.} observed the strong paramagnetic effect 
by measuring the field-angle dependence of the upper critical field. They also showed that the 
paramagnetic effect is suppressed in a few-layer system $M \leq 3$. 
Their experimental results are consistent with our model; 
the paramagnetic effect is suppressed with decreasing $M$ because the superconductivity in surface layers, 
$m=1$ and $M$, is substantially protected against the paramagnetic effect 
due to the RSOC.~\cite{JPSJ.81.034702,Goh} 
Indeed, the upper critical field of the trilayer system is larger than that of the five-layer system near $T=T_{\rm c0}$ 
(see Figs.~6 and 7). 
Note that the upper critical field is dominantly determined from the paramagnetic effect even when 
the orbital effect competes with the RSOC (see the discussion in Sec.~IIIB).

Considering the consistency between our calculation and experiments for CeCoIn$_5$/YbCoIn$_5$ at low magnetic fields, 
it is expected that the PDW state may be realized in the artificial superlattice with $M \simeq 5$ 
at high magnetic fields. 
However, any indication of the presence of high-field superconducting phase has not been reported. 
For instance, the kink and the upturn of the upper critical field shown in our calculations 
have not been observed.~\cite{NatPhys.7.849,Goh,shimozawa} 
The high-field phase may have been missed  because the measurement has not been carried out 
in the low-temperature region. 
On the other hand, the discrepancy may be attributed to the two ingredients 
that are not taken into account in our model. 
First, it is expected that disorders suppress the high-field superconducting phase, as the FFLO state is 
suppressed.~\cite{Matsuda_FFLO} 
The artificial superlattice CeCoIn$_5$/YbCoIn$_5$ indeed contains substantial disorders. 
Second, the number of spacer layers $N= 4 \sim 6$ may not be large enough to eliminate the 
coupling between superconducting multilayers. Inter-multilayer coupling is harmful for the PDW state, and thus 
it should be decreased by increasing the number of spacer layers.

\section{Tricolor Superlattice}

Next, we discuss the tricolor superlattice illustrated in Fig.~\ref{fig1}(b). 
As we have shown in Sec.~IIIB, the orbital effect is controlled by the spacing between the outermost layers. 
Thus, the PDW state may be stabilized in a tricolor superlattice by intercalating the spacer layers into the 
superconducting layers. 
Then, the spacing of neighboring superconducting layers is multiplied to $d=(m_{\rm d}+1)c$ 
with $m_{\rm d}$ being the number of intercalated spacer layers (green open circles in Fig.~\ref{fig1}(b)). 
We assume here that the inter-layer spacing between the spacer layer and the superconducting layer is $c$.  
A similar situation in the bilayer $\delta$-doped SrTiO$_3$~\cite{Inoue} has been realized, and 
superconductivity in the bilayer interface LaAlO$_3$/SrTiO$_3$/LaAlO$_3$ has been studied~\cite{Nakosai}.

Multiplying the inter-layer spacing by $(m_{\rm d} +1)$ is equivalent to increasing the in-plane hopping 
to $(m_{\rm d} +1)t$  while keeping the ratio, $t_\perp/t$, $\mu/t$, $\alpha/t$, and $V_{\rm s}/t$. 
For instance, we obtain the same results for the binary superlattice with $t=80$ meV and for the tricolor superlattice 
with $t=20$ meV and $m_{\rm d}=3$. Thus, the PDW state may be stabilized in the tricolor bilayer superlattice. 
However, in reality, the inter-layer hopping $t_\perp$ between the nearest-neighbor superconducting layers 
is significantly decreased by intercalating a spacer layer. 
For instance, we obtain $t_\perp \sim t_{\rm ns}^2/(E_{\rm s}-E_{\rm n})$ in the presence of a single spacer layer ($m_{\rm d}=1$), 
where $t_{\rm ns}$ is the hopping integral between the superconducting layer and the spacer layer 
and $E_{\rm s}-E_{\rm n}$ is the potential difference between these layers. 
Therefore, the stability of the PDW state in the tricolor superlattice should be examined by investigating the 
superconducting state for small $t_\perp$.

\begin{figure}[htbp]
\begin{center}
\includegraphics[width=60mm]{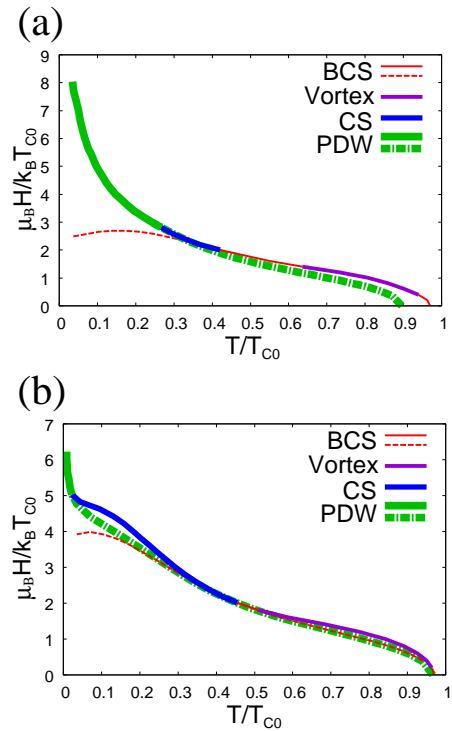}
\caption{(Color online) Transition temperatures of the BCS, CS, vortex, and PDW states 
in the tricolor superlattice with $m_{\rm d} = 3$, $t=20$ meV, $\mu/t=2$, $\alpha/t=0.3$, and $V_{\rm s}/t=1.5$. 
(a) $t_\perp/t=0.05$ and (b) $t_\perp/t=0.01$.
The temperature is normalized by $T_{\rm c0}=0.0124 t/k_{\rm B}$.  
} 
\label{fig10}
\end{center}
\end{figure}

We study here the tricolor bilayer superlattice for simplicity. 
The in-plane hopping is set to $t=20$ meV by considering the heavy-fermion superlattice. 
As we discussed above, we obtain the same phase diagram as Fig.~\ref{fig5} when $m_{\rm d}=3$ and $t_\perp/t =0.1$. 
On the other hand, Figs.~\ref{fig10}(a) and 10(b) show the phase diagram for $t_\perp/t =0.05$ and $t_\perp/t =0.01$, respectively. 
Since the inter-layer Josephson coupling decreases with $t_\perp/t$, 
the uniform states, namely, the BCS and PDW states, are suppressed. 
It is shown that the PDW state is stable at high magnetic fields, 
but the transition temperature of the PDW state is significantly decreased for $t_\perp/t =0.01$. 
The PDW state will be furthermore suppressed by further decreasing $t_\perp/t$. 
Thus, it may be hard to realize the PDW state in a tricolor superlattice by intercalating many spacer layers.

It should be noticed that Fig.~\ref{fig10}(b) is similar to the phase diagram of the binary five-layer superlattice 
(Fig.~\ref{fig7}).  
We now understand that the reduced transition temperature of the PDW state in the five-layer system is 
due to the reduced Josephson coupling between the outermost layers. 
In other words, the superconducting inner layers play a role of the spacer layers. 
Indeed, the PDW state is mainly induced by the outermost layers where the superconductivity is protected against 
the paramagnetic effect by the RSOC. For example, we obtain the layer-dependent order parameter 
$(\Delta_1, \Delta_2, \Delta_3, \Delta_4, \Delta_5) \simeq \Delta (1, 0.18, 0, -0.18, -1)$ in the five-layer PDW state.

Finally, we discuss the superconductivity in the bilayer $\delta$-doped SrTiO$_3$~\cite{Inoue} 
and the bilayer interface LaAlO$_3$/SrTiO$_3$/LaAlO$_3$.~\cite{Nakosai} 
The Fermi velocity estimated for a carrier density $n \sim 10^{14}$ cm$^{-2}$ on the basis of the 
three-orbital tight binding model~\cite{Nakamura2013} is approximately twice as large as that in our model. 
Since the lattice constant along the $c$-axis $c = 3.9$ \AA $\,$ is nearly half of CeCoIn$_5$, 
the orbital effect in the SrTiO$_3$ heterostructures is comparable to our model. 
Thus, the bilayer system sandwiching three non-superconducting layers may be a platform of the odd-parity PDW state. 
Then, the inter-layer coupling between superconducting layers may be small, and therefore, the PDW state may 
appear in the low temperature region, as shown in Fig.~10(b). The kink in the upper critical field would be 
a signature of the PDW state.

\section{Topological Superconductivity}

Topologically non-trivial insulators and SCs have evolved into one of the major research topics 
of modern condensed matter physics recently.~\cite{Qi-Zhang2011,Tanaka2012} 
In particular, topological superconductivity attracts a great deal of attention 
since the Majorana state satisfying the non-Abelian statistics appears 
at the edge and dislocations.~\cite{Read-Green,Ivanov,Kitaev2001} 
In addition to the ``strong'' topological phases classified based on the topological Periodic Table,~\cite{Schnyder,Kitaev2009}  
theories on the symmetry-protected topological superconducting phases (topological crystalline superconductivity) 
have developed recently.~\cite{Zhang,Chiu2013,Morimoto,Shiozaki,Chiu2014}  
In the {\it c}-axis magnetic field the odd-parity PDW state is a topological crystalline SC 
protected by mirror symmetry.~\cite{Yoshida2015} The mirror symmetry along the {\it ab} plane protects the 
topological invariant, that is mirror Chern number~\cite{Ueno-Sato} of symmetry class $D$. 
We obtain the finite mirror Chern number $\nu(\pm i) =\mp 1$, marking the topologically non-trivial properties 
of superconductivity, when the number of superconducting layers $M$ is odd.~\cite{Yoshida2015} 
On the other hand, the mirror Chern number is no longer a topological invariant, when the magnetic field is 
tilted from the $c$ axis.

In this section we show that the PDW state may belong to another kind of topological crystalline SC 
when the magnetic field is applied along the {\it a} or {\it b} axis. 
We demonstrate the topologically non-trivial properties on the basis of the BdG Hamiltonian  
\begin{eqnarray}
{\cal H}_{\rm BdG} &=& \sum_{{\bm k},s,m} \xi({\bm k}+{\bm p}_m) c_{{\bm k}sm}^\dagger c_{{\bm k}sm} 
+t_\perp \sum_{{\bm k},s,\langle m,m'\rangle} c^\dagger_{{\bm k}sm}c_{{\bm k}sm'} 
\nonumber \\
&& - \sum_{{\bm k},s,s',m} \mu_{\rm B} {\bm H} \cdot {\bm \sigma}_{ss'} c_{{\bm k}sm}^\dagger c_{{\bm k}s'm} \nonumber \\ 
&&+\sum_{{\bm k},s,s',m} \alpha_m {\bm g}({\bm k}+{\bm p}_m) \cdot {\bm \sigma}_{ss'}c^\dagger_{{\bm k}sm}c_{{\bm k}s'm} 
\nonumber \\
&&+ \frac{1}{2} \sum_{{\bm k},s,s',m} \left[\Delta_{ss'm}({\bm k}) c^\dagger_{{\bm k}sm}c^\dagger_{-{\bm k}s'm} + {\rm h.c}  \right], 
\label{eq:BdG}
\end{eqnarray}
where $\hat{\Delta}_{m}({\bm k}) \equiv (\Delta_{ss'm}({\bm k})) 
= \left[\psi_m + {\bm d}_m({\bm k}) \cdot {\bm \sigma} \right] i \sigma_y $ 
describes the layer-dependent order parameter of superconductivity.~\cite{Sigrist-Ueda}  
Although the purely $s$-wave superconductivity is considered in Sec.~III and IV, the $p$-wave component is admixed 
through the layer-dependent RSOC by the local violation of inversion symmetry.~\cite{Yoshida2014}  
The layer dependence of order parameter is obtained as 
\begin{eqnarray}
&& \hspace{-5mm}
\left(\psi_1, \psi_2, \psi_3, \psi_4, \psi_5\right) = \left(\psi_{\rm out}, \psi_{\rm in}, 0, -\psi_{\rm in}, -\psi_{\rm out}\right), 
\\
&& \hspace{-5mm}
\left({\bm d}_1({\bm k}), {\bm d}_2({\bm k}), {\bm d}_3({\bm k}), {\bm d}_4({\bm k}), {\bm d}_5({\bm k})\right) 
\nonumber \\ && \hspace{25mm}
= \left(d_{\rm out}, d_{\rm in}, d_{\rm in}', d_{\rm in}, d_{\rm out}\right) \,{\bm g}({\bm k}),
\end{eqnarray}
%
in the five-layer PDW state. 
The BdG Hamiltonian is represented in Nambu space 
  \begin{eqnarray}
&&    {\cal H}_{\rm BdG} = \frac{1}{2} \sum_{{\bm k}} \hat{c}^\dagger \hat{H}_{\rm BdG}({\bm k}) \hat{c},
\\ && 
\hat{H}_{\rm BdG}({\bm k}) =
\left(
\begin{array}{cc}
\hat{H}_0({\bm k}) & \hat{\Delta}({\bm k}) \\
\hat{\Delta}^\dagger({\bm k}) & - \hat{H}_0(-{\bm k})^{T} 
\end{array}
\right),  
    \label{Nambu}
  \end{eqnarray}
where $\hat{H_0}({\bm k})$ is the Hamiltonian in the normal state and $\hat{\Delta}({\bm k})$ is the order parameter. 

Although the mirror symmetry with respect to the {\it ab} plane ${\cal M}_{ab}$ is broken 
in the in-plane magnetic field, 
the magnetic mirror symmetry ${\cal T'}= {\cal T M}_{ca}$ (${\cal T M}_{bc}$) is preserved 
in the {\it a}-axis ({\it b}-axis) magnetic field. 
For instance, in the magnetic field along the {\it a} axis the BdG Hamiltonian is invariant 
under magnetic mirror symmetry 
\begin{eqnarray}
&& \hspace{-5mm} {\cal T'} \hat{H}_{\rm BdG}(k_x, k_y) {\cal T'}^{\,\dagger} = \hat{H}_{\rm BdG}(-k_x, k_y), 
\end{eqnarray}
where ${\cal M}_{ca} = i \sigma_y$ is the mirror reflection operator and ${\cal T} = i \sigma_y {\cal K}$ is 
the time-reversal operator with ${\cal K}$ being the complex conjugate operator. 
Combining with the particle-hole symmetry 
${\cal C} \hat{H}_{\rm BdG}({\bm k}){\cal C}^\dagger = -\hat{H}_{\rm BdG}(-{\bm k})$, where 
${\cal C}= \tau_x {\cal K}$ and $\tau_x$ is the Pauli matrix in the particle-hole space, 
we can define the mirror chiral symmetry 
$\Gamma \hat{H}_{\rm BdG}(k_x,k_y)\Gamma^\dagger = -\hat{H}_{\rm BdG}(k_x,-k_y)$ with 
$\Gamma = -{\cal C T'} = \tau_x$. 
Thus, the BdG Hamiltonian satisfies the chiral symmetry 
\begin{eqnarray}
&& \{ \Gamma , \hat{H}_{\rm BdG}({\bm k})\}=0 
\label{chiral}
\end{eqnarray} 
at $k_y =0$ and $k_y = \pi/a$. The chiral symmetry ensures that the one-dimensional winding number
\begin{eqnarray}
\label{winding}
{\cal \omega}_{k_y} = \frac{1}{4\pi i} \int_{-\frac{\pi}{a}}^{\frac{\pi}{a}} dk_x {\rm Tr}
\left[\hat{q}({\bm k})^{-1}\partial_{k_x}\hat{q}({\bm k}) 
- \hat{q}^\dagger({\bm k})^{-1}\partial_{k_x}\hat{q}^\dagger({\bm k}) \right] 
\nonumber \\
\end{eqnarray}
is a topological invariant~\cite{Sato-Fujimoto2009,Sato2011,Yada2011,Schnyder2011,Mizushima_3He,Tsutsumi_UPt3} 
when a finite gap is open at $k_y =0$ and $k_y = \pi/a$. 
The $2M \times 2M$ matrix $\hat{q}({\bm k})$ is obtained by carrying out the unitary transformation 
\begin{eqnarray}
&& 
U \hat{H}_{\rm BdG}({\bm k}) U^\dagger =
\left(
\begin{array}{cc}
0 & \hat{q}({\bm k}) \\
\hat{q}^\dagger({\bm k}) & 0 \\
\end{array}
\right). 
\label{off-diagonal}
\end{eqnarray}

When we regard the magnetic mirror symmetry ${\cal T'}$ as pseudo-time-reversal symmetry,~\cite{Tewari,Wong-Law} 
the one-dimensional Hamiltonian $\hat{H}_{\rm 1D}^{k_y =0}(k_x) = \hat{H}_{\rm BdG}(k_x,0)$ and 
$\hat{H}_{\rm 1D}^{k_y =\pi/a}(k_x) = \hat{H}_{\rm BdG}(k_x,\pi/a)$ belong to the symmetry class $BDI$ 
because ${\cal T'}^2 =+1$.~\cite{Schnyder,Kitaev2009} 
Thus, we can define the integer topological numbers of the $BDI$ class 
\begin{eqnarray}
\label{BDI}
&& \hspace{-5mm}
\nu^{\rm BDI}_0 = \frac{1}{\pi i} \int_{0}^{\frac{\pi}{a}} dk_x {\rm Tr}
\left[\hat{q}(k_x,0)^{-1}\partial_{k_x}\hat{q}(k_x,0) \right],  
\\ && \hspace{-5mm}
\nu^{\rm BDI}_{\pi/a} = \frac{1}{\pi i} \int_{0}^{\frac{\pi}{a}} dk_x {\rm Tr}
\left[\hat{q}(k_x,\pi/a)^{-1}\partial_{k_x}\hat{q}(k_x,\pi/a) \right]. 
\label{BDI2}
\end{eqnarray}
Indeed, these winding numbers are equivalent to Eq.~(\ref{winding}), namely, $\nu^{\rm BDI}_{0,\pi/a} = {\cal \omega}_{0,\pi/a}$. 
The pseudo-time-reversal symmetry considered here has been used for the definition of the integer topological number 
in one-dimensional semiconductor nanowires~\cite{Tewari} and quasi-one-dimensional $d$-wave superconductors.~\cite{Wong-Law} 
The magnetic mirror symmetry is the physical origin of this ``hidden'' time-reversal symmetry. 
The difference of two winding numbers, $\nu^{\rm BDI}_{0} - \nu^{\rm BDI}_{\pi/a}$, is the strong index of 2D topological 
crystalline SCs protected by the magnetic mirror symmetry.~\cite{Shiozaki}

\begin{figure}[htbp]
\begin{center}
\includegraphics[width=80mm]{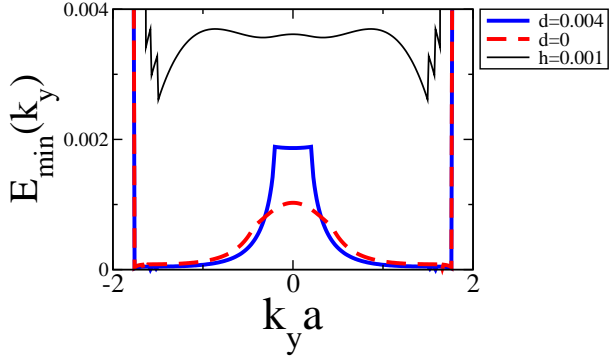}
\caption{(Color online) 
Superconducting gap at each $k_y$ in the five-layer PDW state, defined by 
$E_{\rm min}(k_y) = {\rm Min}_{i,k_x} |E_i({\bm k})|$. 
We assume $t=20$ meV, $t_\perp/t=0.1$, $\mu/t=-2$, $\alpha/t=0.3$, and $\mu_{\rm B}H/t =0.04$ is the magnetic field 
along the {\it a} axis. 
The layer-dependent order parameters are chosen as $\psi_{\rm out}=0.02$ and  $\psi_{\rm in}=0.0036$ consistent with 
the results in Sec.~IIIB (Fig.~7). We take into account a small $p$-wave component 
$d \equiv d_{\rm out}=d_{\rm in}=d_{\rm in}'=0.004$ induced by the RSOC~\cite{Yoshida2014} (thick solid line), 
while we obtain the dashed line for $d=0$. We also show the gap for $d=0.004$ and $\mu_{\rm B}H/t =0.001$ for a comparison 
(thin solid line). Bulk gap opens in the low magnetic field region, but it closes in the high magnetic field region 
where the PDW state is stable. The superconducting gap at $k_y = 0$ is finite even in the high magnetic field region. 
} 
\label{fig11}
\end{center}
\end{figure}

We now discuss the superconducting gap. 
Figure~\ref{fig11} shows the gap of the single-particle excitation spectra in the five-layer PDW state 
for each $k_y$, which is defined as $E_{\rm min}(k_y) = {\rm Min}_{i,k_x} |E_i({\bm k})|$, 
with $E_i({\bm k})$ being eigenvalues of the BdG Hamiltonian $\hat{H}_{\rm BdG}({\bm k})$. 
We here assume $\mu/t=-2$ so that the Fermi surface encloses the $\Gamma$ point (${\bm k}={\bm 0}$). 
Since there is no Fermi surface along $k_y =\pi/a$, the winding number is trivial, $\nu^{\rm BDI}_{\pi/a} = 0$.
Therefore, we focus on $\nu^{\rm BDI}_{0}$. 
The superconducting gap is finite at $k_y =0$, ensuring the topological protection of the winding number $\nu^{\rm BDI}_{0}$. 
At low magnetic fields, the superconducting gap is finite in the whole Brillouin zone (thin solid line in Fig.~11), 
and thus, $\nu^{\rm BDI}_{0}$ is the strong topological index. 
Although the gap at finite $k_y$ is closed at high magnetic fields owing to the paramagnetic effect, 
the winding number is regarded as a topological number of an effective one-dimensional Hamiltonian.

\begin{figure}[htbp]
\begin{center}
\includegraphics[width=68mm]{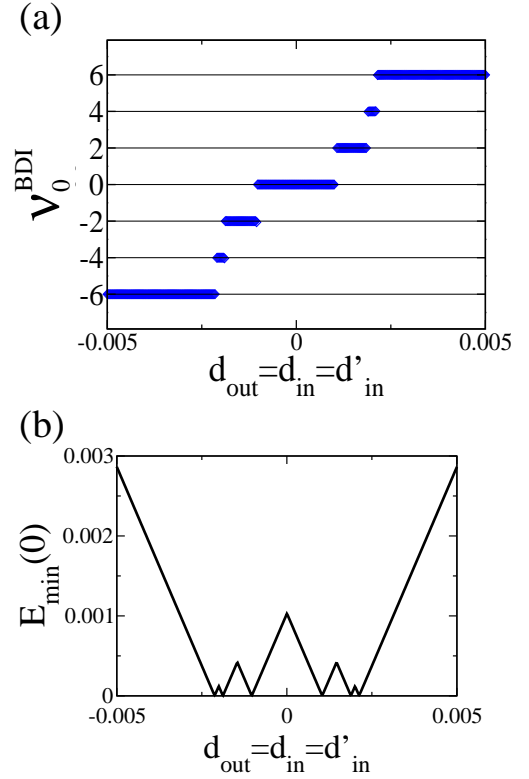}
\caption{(Color online)
(a) Winding number and (b) superconducting gap at $k_y =0$ as a function of the $p$-wave component 
$d \equiv d_{\rm out} = d_{\rm in} = d_{\rm in}'$. 
The other parameters are the same as those in Fig.~\ref{fig11}.
} 
\label{fig12}
\end{center}
\end{figure}

As we show in Fig.~\ref{fig12}(a), $\nu^{\rm BDI}_{0}$ discretely changes upon increasing the $p$-wave 
component in the order parameter, $d \equiv d_{\rm out} = d_{\rm in} = d_{\rm in}'$. 
The superconducting gap is closed for special values of $d$ where the winding number jumps (Fig.~\ref{fig12}(b)). 
We here ignore the layer dependence of the $p$-wave component for simplicity. 
This assumption has been justified by the BdG equation, which shows the nearly 
layer-independent $p$-wave component in the PDW state.~\cite{Yoshida2014}

\begin{figure}[htbp]
\begin{center}
\includegraphics[width=70mm]{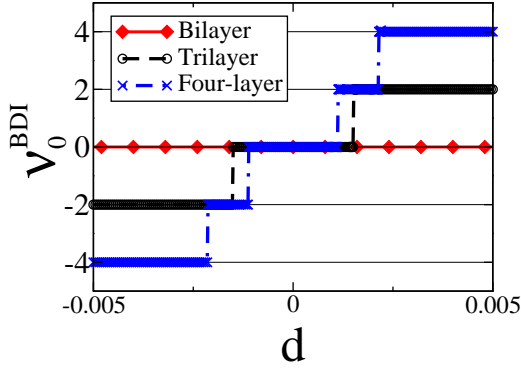}
\caption{(Color online)
Winding number of the bilayer, trilayer, and four-layer PDW state. 
We assume a layer-independent $p$-wave component $d \equiv d_{\rm out} = d_{\rm in}$ as in Fig.~\ref{fig12}. 
The $s$-wave component has a layer-dependence 
$\left(\psi_1, \psi_2 \right) = \left(\psi_{\rm out}, -\psi_{\rm out} \right)$, 
$\left(\psi_1, \psi_2, \psi_3 \right) = \left(\psi_{\rm out}, 0, -\psi_{\rm out} \right)$, and 
$\left(\psi_1, \psi_2, \psi_3, \psi_4 \right) = \left(\psi_{\rm out}, \psi_{\rm in}, -\psi_{\rm in}, -\psi_{\rm out} \right)$.  
The parameters are the same as those in Fig.~\ref{fig11}.
} 
\label{fig13}
\end{center}
\end{figure}

Switching on a small $p$-wave component $|d| \geq 0.05 \psi_{\rm out}$, we obtain a finite winding number indicating 
the topologically non-trivial properties [see Fig.~\ref{fig12}(a)]. This is in sharp contrast to the 2D Rashba SC, 
where the $p$-wave component overwhelming the $s$-wave component is required for topological 
superconductivity.~\cite{Sato-Fujimoto2009,Tanaka} This condition is hardly realized in real materials. 
On the other hand, a small $p$-wave component induced by the RSOC causes the PDW state 
to be topologically non-trivial, because the $s$-wave component of the order parameter is small 
on inner layers when $M \ge 3$. 
In Fig.~\ref{fig13}, we show the winding number of the bilayer, trilayer, and four-layer systems in the PDW state. 
Although the bilayer system ($M=2$) is trivial, the topologically non-trivial superconducting state is 
induced by a small $p$-wave component for $M \ge 3$. 
According to the random-phase-approximation (RPA) analysis of the three-dimensional 
Rashba-Hubbard model, the induced spin-triplet component is approximately 20\% of 
the spin-singlet component for a moderate RSOC.~\cite{Yanase2007,Yanase2008} 
Thus, the $p$-wave component is likely to be large enough to realize the topological crystalline 
superconductivity protected by magnetic mirror symmetry for $M \ge 3$. 
Note that these conditions are different from those for the topological superconductivity in the magnetic field 
along the {\it c} axis. 
The PDW state is a topological crystalline superconductor in the {\it c}-axis magnetic field 
when the number of superconducting layers $M$ is odd.~\cite{Yoshida2015} Then, the $p$-wave component 
is not required.

\begin{figure}[htbp]
\begin{center}
\includegraphics[width=65mm]{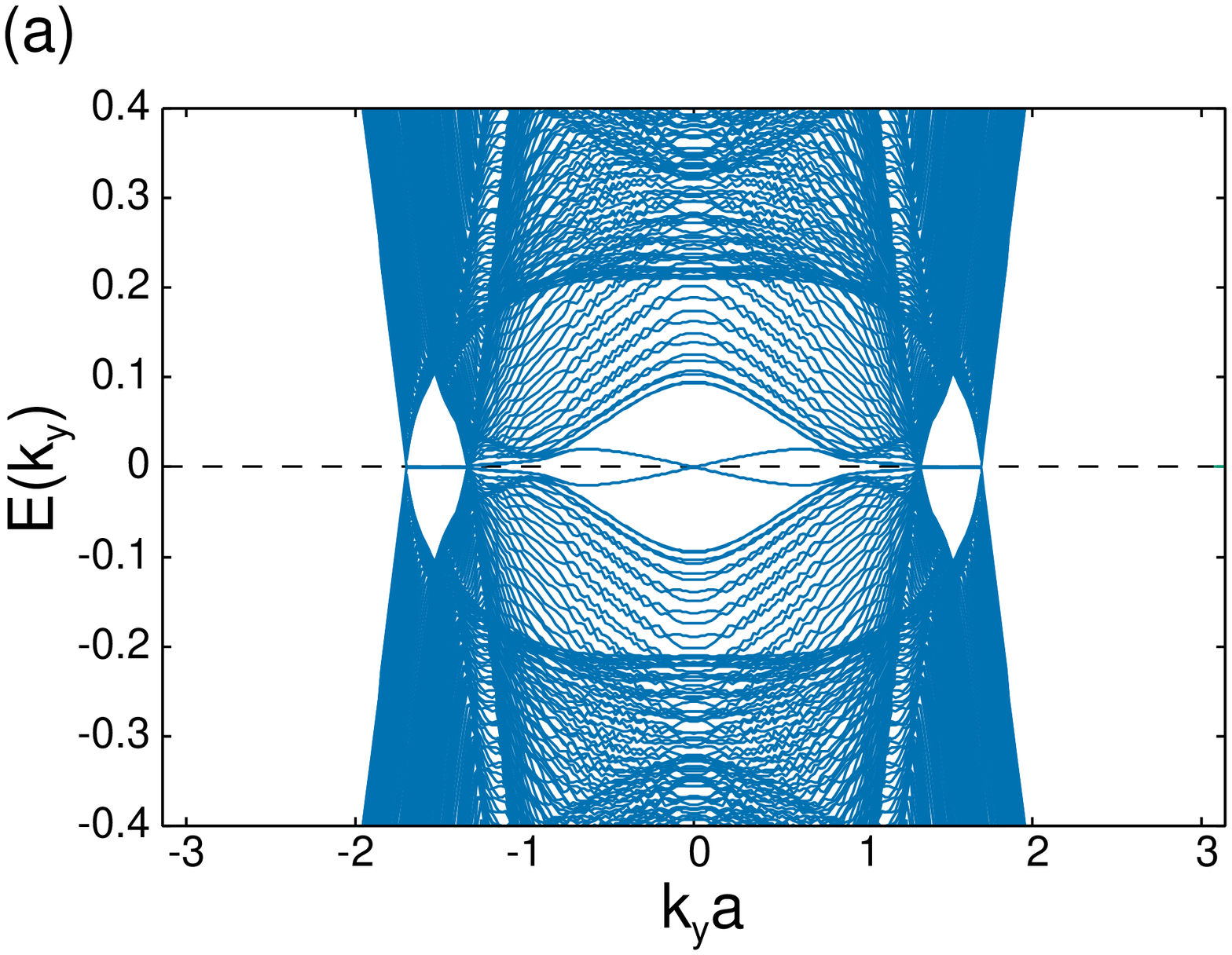}
\includegraphics[width=65mm]{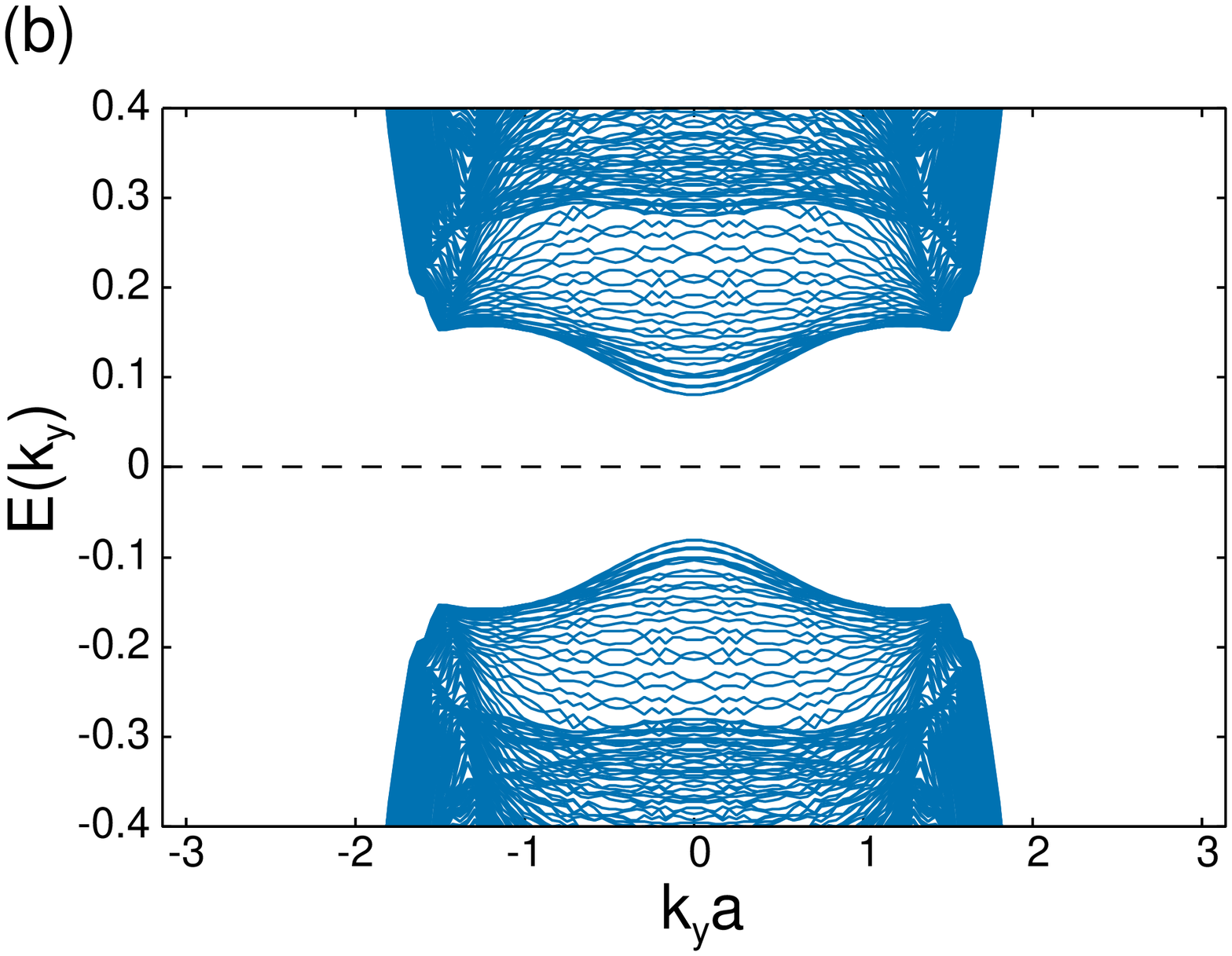}
\caption{(Color online) Energy spectra in the trilayer PDW state with open boundaries along the {\it a} axis. 
We consider the magnetic field along the {\it a}-axis $\mu_{\rm B}H/t=0.3$ (a) and the {\it b} axis $\mu_{\rm B}H/t=0.1$ (b). 
We assume $t_\perp/t=0.1$, $\mu/t=-2$, $\alpha/t=0.3$, and choose the $s$-wave order parameter $\psi_{\rm out}=0.5$ and 
the $p$-wave order parameter $d=0.2$. The orbital effect is neglected  for simplicity. 
} 
\label{fig14}
\end{center}
\end{figure}

A non-trivial winding number may ensure the presence of the Majorana edge state 
according to the bulk-edge correspondence. 
When the magnetic field is applied along the {\it a} axis, the magnetic mirror symmetry ${\cal T M}_{ca}$ is preserved 
at the edge perpendicular to the [100]-axis [(100) edge]. Therefore, the winding number protected by this symmetry 
corresponds to the number of zero-energy edge states according to the index theorem.~\cite{Sato2011} 
Indeed, we show the Majorana edge states in Fig.~\ref{fig14}. The trilayer PDW state 
with a large superconducting gap is considered for simplicity of numerical calculation. 
The energy spectrum is calculated in the open boundary condition 
along the {\it a} axis. The layer-dependent order parameters are 
$\left(\psi_1, \psi_2, \psi_3\right) = \left(1, 0, -1\right)\psi_{\rm out}$ and 
$\left({\bm d}_1({\bm k}), {\bm d}_2({\bm k}), {\bm d}_3({\bm k})\right) = \left(1,1,1\right) d \, {\bm g}({\bm k})$. 
We see the two Majorana modes around $k_y =0$ (Fig.~\ref{fig14}(a)), 
when we assume a small $p$-wave component leading to 
the winding number $\nu^{\rm BDI}_{0}=2$. The Majorana states have a linear dispersion 
since the chiral symmetry defined in Eq.~(\ref{chiral}) is not preserved at $0 < |k_y| < \pi/a$. 
Because another pseudo-time-reversal symmetry, 
${\cal T''} \hat{H}_{\rm BdG}({\bm k}) {\cal T''}^\dagger  = \hat{H}_{\rm BdG}(-{\bm k})$ 
with ${\cal T''}={\cal T}{\cal M}_{ab}$, 
is preserved, the two Majorana states form ``Kramers pairs''. 

Figure~\ref{fig14}(a) also shows the zero-energy flat band at $|k_y a| = 1.35$-$1.7$. 
This mode is specified by another winding number $\omega'_{k_y}$ protected by 
the pseudo-time-reversal symmetry ${\cal T''}$.  
The winding number $\omega'_{k_y}$ is defined at all $k_y$, and we obtain $\omega'_{k_y} =-1$ at $k_y$ where the flat band appears. 
Hence, the zero-energy flat band does not have any degeneracy.

Finally, we comment on the anisotropic response to the external magnetic field. 
The magnetic mirror symmetry ${\cal T M}_{ca}$ is broken when we apply the magnetic field 
along the {\it b} axis. Then, the zero-energy Majorana states disappear at the (100) edge, 
as expected [Fig.~\ref{fig14}(b)]. 
This field angle dependence is attributed to the Ising character of the Majorana state. 
Similarly, the Majorana mode appears (disappears) at the (010) edge in the magnetic field along the {\it b} axis 
({\it a} axis), because the mirror symmetry along the {\it bc} plane ${\cal M}_{bc}$ is preserved at the edge.

\section{Summary and Discussion}

In this paper we studied 2D multilayer SCs influenced by the layer-dependent RSOC. 
We showed that the odd-parity PDW state is stabilized by the competing spin-orbit coupling and the orbital effect 
in the magnetic field along the 2D conducting plane. We also showed that the PDW state is a topological crystalline 
SC protected by the magnetic mirror symmetry when a small $p$-wave component is induced by the RSOC.
The Majorana state has been demonstrated at the (100) edge [(010) edge] in the magnetic field 
along the {\it a} axis ({\it b} axis). 

Our finding paves the way toward realizing odd-parity superconductivity without a considerable pairing interaction 
in the spin-triplet channel. Although spin-triplet superconductivity is hardly stabilized in most SCs 
except for a few exceptions, our proposal provides an alternative way to create odd-parity SC 
by using the sublattice degree of freedom. 

Indeed, recent developments in the technology of artificial heterostructures may enable 
the design of the odd-parity PDW state. 
Superconducting 2D electron systems have been fabricated in the oxide interfaces 
SrTiO$_3$/LaAlO$_3$~\cite{Reyren} and SrTiO$_3$/LaTiO$_3$,~\cite{Biscaras} 
gate-tuned SrTiO$_3$~\cite{Ueno} and MoS$_2$,~\cite{Ye}  
and the heavy-fermion superlattice CeCoIn$_5$/YbCoIn$_5$.~\cite{NatPhys.7.849} 
It has been reported that interfacial (intrinsic) spin-orbit coupling significantly affects 
the superconducting state in SrTiO$_3$ heterostructures~\cite{BenShalom,Caviglia_Rashba,Michaeli,Nakamura2013} 
and CeCoIn$_5$/YbCoIn$_5$~\cite{Goh,shimozawa} (MoS$_2$~\cite{Saito,Lu}). 
Furthermore, the multilayer structure has been artificially controlled in CeCoIn$_5$/YbCoIn$_5$~\cite{shimozawa} 
and $\delta$-doped SrTiO$_3$.~\cite{Inoue} 
Thus, we expect that odd-parity topological superconductivity 
will be created in these systems by tuning the multilayer structure and the magnetic field.

\section*{Acknowledgements} 
The authors are grateful to Y. Iwasa, Y. Matsuda, T. Shibauchi, M. Sigrist, M. Shimozawa, and Y. Tada 
for fruitful discussions. 
This work was supported by the "Topological Quantum Phenomena" (No. 25103711) 
Grant-in Aid for Scientific Research on Innovative Areas from MEXT of Japan, 
and by JSPS KAKENHI Grant Numbers 24740230, 15K05164, 15H05745, and 15H05884 (J-Physics).

\end{document}